# Proximate Kitaev Quantum Spin Liquid Behavior in α-RuCl₃

A. Banerjee[1]*, C.A. Bridges[2], J-Q. Yan[3,4], A.A. Aczel[1], L. Li[4], M.B. Stone[1], G.E. Granroth[5], M.D. Lumsden[1], Y. Yiu[6], J. Knolle[7], D.L. Kovrizhin[7], S. Bhattacharjee[8], R. Moessner[8], D.A. Tennant[9], D.G. Mandrus[3,4], S.E. Nagler[1,10]*

[1]Quantum Condensed Matter Division, Oak Ridge National Laboratory, Oak Ridge, TN, 37830, U.S.A.

[2]Chemical Sciences Division, Oak Ridge National Laboratory, Oak Ridge, TN, 37830, U.S.A.

[3]Materials Science and Technology Division, Oak Ridge National Laboratory, Oak Ridge, TN, 37830, U.S.A.

[4]Department of Materials Science and Engineering, University of Tennessee, Knoxville, TN, 37996, U.S.A.

[5]Neutron Data Analysis & Visualization Division, Oak Ridge National Laboratory, Oak Ridge – TN 37830, U.S.A.

[6]Department of Physics, University of Tennessee, Knoxville, TN, 37996, U.S.A.

[7]Department of Physics, Cavendish Laboratory, JJ Thomson Avenue, Cambridge CB3 0HE, U.K.

[8]Max Planck Institute for the Physics of Complex Systems, D-01187 Dresden, Germany.

[9]Neutron Sciences Directorate, Oak Ridge National Laboratory, Oak Ridge, TN, 37830, U.S.A.

[10]Bredesen Center, University of Tennessee, Knoxville, TN, 37966, U.S.A.

**Topological states of matter such as quantum spin liquids (QSLs) are of great interest because of their remarkable predicted properties including protection of quantum information and the emergence of Majorana fermions. Such QSLs, however, have proven difficult to identify experimentally. The most promising approach is to study their exotic nature via the wave-vector and intensity dependence of their dynamical response in neutron scattering. A major search has centered on iridate materials which are proposed to realize the celebrated Kitaev model on a honeycomb lattice – a prototypical topological QSL system in two dimensions (2D). The difficulties of iridium for neutron measurements have, however, impeded progress significantly. Here we provide experimental evidence that a material based on ruthenium, α-RuCl₃ realizes the same Kitaev physics but is highly amenable to neutron investigation. Our measurements confirm the requisite strong spin-orbit coupling, and a low temperature**

**magnetic order that matches the predicted phase proximate to the QSL. We also show that stacking faults, inherent to the highly 2D nature of the material, readily explain some puzzling results to date. Measurements of the dynamical response functions, especially at energies and temperatures above that where interlayer effects are manifest, are naturally accounted for in terms of deconfinement physics expected for QSLs. Via a comparison to the recently calculated dynamics from gauge flux excitations and Majorana fermions of the pure Kitaev model we propose α-RuCl$_3$ as the prime candidate for experimental realization of fractionalized Kitaev physics.**

Exotic physics associated with frustrated quantum magnets is an enduring theme in condensed matter research. The formation of quantum spin liquids (QSL) in such systems can give rise to topological states of matter with fractional excitations[1,2,3,4]. The realization of this physics in real materials is an exciting prospect that may provide a path to a robust quantum computing technology[4,5]. Fractional excitations in the form of pairs of S=1/2 spinons are observed in quasi-one dimensional materials containing S=1/2 Heisenberg antiferromagnetic chains[6]. Recent evidence for the 2D QSL state, in the form of possible spinon excitations, has been found in quantum antiferromagnets on triangular[3] and Kagome[7] lattices. On the theory side, the exactly solvable Kitaev model on the honeycomb lattice[8] represents a class of 2D QSL that supports two different emergent fractionalized excitations: Majorana fermions and gauge fluxes[9,10]. This work is an experimental search for evidence of these excitations.

The Kitaev model consists of a set of spin-1/2 moments $\{\vec{S}_i\}$ arrayed on a honeycomb lattice. The Kitaev couplings, of strength $K$ in eqn. (1) are highly anisotropic with a different spin component interacting for each of the three bonds of the honeycomb lattice. In actual materials a Heisenberg interaction ($J$) is also generally expected to be present, giving rise to the Heisenberg-Kitaev (H-K) Hamiltonian given by[11].

$$\mathcal{H} = \sum_{i,j}\left(K S_i^m S_j^m + J \vec{S}_i \cdot \vec{S}_j\right) \qquad \text{eqn. (1)}$$

where, for example, m is the component of the spin directed along the bond connecting spins (i,j). The QSL phase of the pure Kitaev model ($J$=0), for both ferro and antiferromagnetic $K$, is stable for relatively small Heisenberg perturbations.

Remarkably the Hamiltonian (1) has been proposed to accurately describe octahedrally-coordinated magnetic systems, Fig. 1, with dominant spin-orbit coupling[11]. The focus to date has centered largely on Ir$^{4+}$ compounds[12-16], however attempts to measure the dynamical response[14] via inelastic neutron scattering (INS) have met with limited success, due to the unfavorable magnetic form factor and strong absorption cross-section of the Ir ions. Resonant inelastic x-ray scattering (RIXS) has provided important information concerning higher energy excitations in the iridates[17] but cannot provide the meV energy resolution necessary to provide a robust experimental signature of fractional excitations.

An alternative approach is to explore materials with $Ru^{3+}$ ions[18]. The realization that the material $\alpha$-$RuCl_3$[19,20] also has the requisite honeycomb lattice and strong spin-orbit coupling has stimulated a groundswell of recent investigations[21-27]. Whilst these studies lend support to the material as a potential Kitaev material, conflicting results centering on the low temperature magnetic properties have hindered progress. To resolve this we undertake a comprehensive evaluation of the magnetic and spin orbit properties of $\alpha$-$RuCl_3$, and further measure the dynamical response establishing this as a material proximate to the widely searched for quantum spin liquid.

We begin by investigating the crystal and magnetic structure of $\alpha$-$RuCl_3$. The layered structure of the material is shown in Fig. 1(a). Figures 1 (b) and (c) show the ABCABC stacking arrangement of the layers expected in the trigonal structure. That the layers are very weakly bonded to each other, similar to graphite, is demonstrated by the lattice specific heat (shown for a powder in Fig. 1(d)). This displays a tell-tale $T^2$ behavior characteristic of highly 2D bonded systems rather than the usual $T^3$ observed in conventional 3D solids[28]. Since the 2D layers are weakly coupled the inter-layer magnetic exchanges will also be very weak. In addition stacking faults will be formed easily[23] and significant regions of the sample can crystallize in alternative stacking structures, for example ABAB.

Neutron diffraction (see methods and Supplementary Information (SI)) shows low temperature magnetic order. The temperature dependence of the strongest magnetic powder peak, with $T_N \approx 14$ K, is shown in figure 1(e). Figure 1(f) shows the temperature dependence of magnetic peaks in a 22.5 mg single crystal, revealing two ordered phases. The first, which orders below $T_N \approx 14$ K, is characterized by a wavevector of $\mathbf{q_1}$ = (1/2 0 3/2) (indexed according to the trigonal structure), whilst the other phase ($\mathbf{q_2}$ = (1/2, 0 1)) orders below 8 K. These temperatures correspond precisely to anomalies observed in the specific heat and magnetic susceptibility[23,24,27]. This is readily explained as the observed L =3/2 phase corresponds naturally to a stacking order of ABAB type along the c-axis, and the L = 1 corresponds to ABCABC stacking. Indeed, the difference in 3D transitions is a residual effect of different interlayer bonding influencing the ordering. Further, a comparison of intensities at (1/2 0 L) with (3/2 0 L)[15] shows both phases share identical zig-zag (ZZ) spin ordering in the honeycomb layers; a phase of the H-K model adjacent to the spin liquid. By calibrating to structural Bragg peaks the ordered moments are measured to be exceptionally low, with an upper bound of $\mu$=0.4 $\pm$ 0.1 $\mu_B$. This is at most only 35% of the full moment determined from bulk measurements[23] (see methods) suggesting strong fluctuations consistent with a near liquid-like quantum state in the material.

Having established the structural and magnetic properties of α-RuCl$_3$, we probe the nature of the single ion states to confirm the presence of strong spin-orbit coupling, which is required to generate the Kitaev term *K* in (1). By using 1.5 eV incident neutrons to measure the transition from the Ru$^{3+}$ electronic ground state to its excited state (see methods) the spin orbit coupling λ is extracted. In the octahedral environment shown in Fig. 1 the ground state is a low-spin (J=1/2) state. The next excited state (J=3/2) is separated by 3λ/2. Neutrons can activate it by a spin flip process and the transition is seen in Fig. 2 at 195 ± 11 meV implying that λ ≈ 130 meV. This is close to the expected free-ion value (λ$_{free}$ ≈ 150 meV[18,29]) and recent ab-initio calculations[24]. The J=3/2 state will be split into two Kramers doublets by small distortions of the octahedron[30,31]. The resolution limited line-width suggests that such a splitting is small. In any case, as the higher levels are too energetic to play any role, only the lowest lying doublet needs to be considered and inter-Ru$^{3+}$ couplings then project out into the Kitaev form as required.

The above results indicate that a H-K Hamiltonian (1) can be a good low energy description of the interactions between Ru$^{3+}$ moments. To further elucidate the effective spin Hamiltonian we make measurements of the collective magnetic excitations. Fig. 3 shows data for α-RuCl$_3$ powder collected using the SEQUOIA spectrometer at the Spallation Neutron Source, Oak Ridge National Laboratory. The scattering in the magnetically ordered state is shown in Fig. 3(a) for T = 5 K. Two branches of excitations are clearly visible. The lower branch, M$_1$, is centered near 4 meV and shows a minimum near Q = 0.62 Å$^{-1}$, which notably corresponds to the M point of the honeycomb lattice as expected for a quasi-2D magnetic system (for 3D behavior a wave-vector Q = 0.81 Å$^{-1}$ is anticipated). The white arrow draws attention to the concave shape of the edge of the scattering, which is expected for magnon excitations in a ZZ ordered state[14]. This firmly locates the H-K phase and differentiates from other potential states such as a stripy ground state. Meanwhile a second, higher energy, mode, M$_2$, is centered near 6.5 meV.

Both the modes M$_1$ and M$_2$ have a magnetic origin as identified by their wave-vector and temperature dependence. As well as these, a phonon mode (marked "P") contributes to the scattering at an energy near that of M$_2$ but at higher wave-vectors of Q > 2 Å$^{-1}$. The thermal behavior of these magnetic modes differs significantly from one to the other. Fig. 3(b) shows the scattering at T = 15 K, just above T$_N$. It is seen that M$_1$ softens dramatically and the intensity shifts towards Q = 0. Conversely, M$_2$ is essentially unaffected. Constant Q cuts through the data are displayed in Fig. 3(c). The centers are at the positions indicated by the labeled dashed lines in Fig. 3(a) and 3(b). Comparing cuts (A,B) with (C,D) reinforces the collapse and shift of intensity for M$_1$ above T$_N$. Cut B clearly shows two peaks implying that the density of states sampled by the powder average at T = 5 K has two maxima. The average peak energies determined by fits of the data to Gaussian peaks are given by $E_1$ = 4.1(1) meV and $E_2$ = 6.5(1) meV. Fig. 3(d) shows constant energy cuts integrated over the range [2.5, 3.0] meV, near the lower edge of $M_1$. It is seen that at low temperature $M_1$ is structured with low energy features showing up as peaks in cut E. These are centered at $Q_1$ = 0.62(3) Å$^{-1}$ and $Q_2$ = 1.7(1) Å$^{-1}$. Above $T_N$ this structure disappears, and the broad scattering shifts dramatically to lower Q. Fitting the T = 15 K data (cut F) to a Lorentzian with the center fixed at Q = 0 yields a HWHM of roughly 0.6 Å$^{-1}$, suggesting that above $T_N$ spatial correlations of the spin fluctuations are extremely short ranged and do not extend much if at all beyond the neighboring spins.

To gain insight into the strength and nature of the magnetic couplings we compare the INS data to the solution of (1) using conventional linear spin-wave theory (SWT) for ZZ order[32,33]. In the honeycomb lattice appropriate for $\alpha$-RuCl$_3$, SWT predicts four branches, two of which disperse from zero energy at the M point (½, 0) to doubly degenerate energies $\omega_1 = \sqrt{K(K+J)}$ and $\omega_2 = |J|\sqrt{2}$ respectively at the $\Gamma$ point (0,0)[32]. A large density of states in the form of van Hove singularities is expected near $\omega_1$ and $\omega_2$. Fig. 4 (a) shows the theory and fig. 4(b) the calculated powder averaged neutron scattering. The measurements locate the energies $E_1$ and $E_2$ above and comparison to SWT yields K and J values of (K=7.0, J=-4.6) meV or (K=8.1, J=-2.9) meV depending on whether $\omega_1$ corresponds to $E_1$ or $E_2$. The inset of Fig. 4(d) shows each of these possibilities on the H-K phase diagram[32]. Either way the Kitaev term is stronger and antiferromagnetic, while the Heisenberg term is ferromagnetic; again consistent with ab-initio calculations[25].

Although the calculation reproduces many of the features of the observed dynamical response, crucial *qualitative* differences remain. First, the $M_1$ mode has a clear and significant gap of approx. 1.7 meV near the M point, see Fig. 5, at odds with SWT. While such a gapless spectrum is a known artifact of linear SWT for the H-K model[32], the experimentally observed gap is too large to be accounted for within systematic 1/S corrections. Extending the Hamiltonian to include further terms can lead to a gap forming within SWT. However, calculations of the SW spectrum (see SI) with additional terms in the Hamiltonian (such as $\Gamma$ and/or $\Gamma'$ terms[34–37]), when sufficient to generate the observed gap, show additional features in the powder averaged scattering that are inconsistent with the observations. Within the SW approximation a gap can also be generated by adding an additional Ising-like anisotropy, however at present there is no justification for the presence of such a term in $\alpha$-RuCl$_3$. Second, and more importantly, linear SWT is not compatible with the width or temperature dependence of the upper branch, $M_2$. Fig. 5(b) shows a constant Q cut around the $M_2$ mode and the equivalent SW calculation, broadened by the instrumental energy resolution. $M_2$ is much broader than the SW calculation, as seen clearly on the high energy side of the peak (shaded region). A plot of the $M_2$ intensity as a function of Q is shown in Fig 5(c) for T = 5 K and 15 K. The intensities remain identical *across the Neel transition*, and decrease monotonically as Q is increased.

The SWT is a quantization of harmonic excitations from semi-classical order. As noted above, a gap is expected to arise from quantum effects that are not captured by linear SWT[32]. Moreover, the very low ordered moment observed in $\alpha$-RuCl$_3$ indicates that linear SWT is inapplicable. The mismatched features described above are expected to arise from non-linear dynamical effects and strong quantum fluctuations not captured by a linear SWT. Indeed, we argue that the observed higher energy mode $M_2$ scattering – which because of its short-time scale is least sensitive to 3D couplings – is naturally accounted for in terms of a QSL phase proximate in the H-K phase diagram[38].

This QSL viewpoint has as its starting point the strong quantum limit. It can avail itself of the recently computed exact dynamical structure factor of the pure Kitaev model, in which spin excitations fractionalize into static Ising fluxes and propagating Majorana fermions minimally coupled to a $Z_2$ gauge field[10]. Powder averaged results of the scattering expected are shown for the isotropic antiferromagnetic Kitaev model in Fig. 5(d). Although the QSL is gapless, the structure factor shown in Fig. 5(d) does show a gap because of the heavy gapped $Z_2$ fluxes[10]. This results in a low energy band from 0.125-0.5 *K* with a peak of intensity near the M point in the Brillouin zone. Most interestingly, in addition there appears a second very broad and non-dispersing high energy band centered at an energy that corresponds approximately to the Kitaev exchange scale, *K*. The intensity of the upper band is strongest at Q =0 and decreases with Q.

With the Kitaev interaction dominant it is reasonable to expect that $\alpha$-RuCl$_3$ is proximate to the QSL phase. This leads to a natural interpretation of the $M_2$ mode as having the same origin as the high energy mode of the QSL. Conversely, the lower-energy features are dependent on the details of the phases on either side of the transition. The broad nature of the $M_2$ mode, as seen in the measurements, is then naturally explained in terms of the fractionalized Majorana fermion excitations. Moreover, the Q dependence of the intensity of the $M_2$ mode intensity strikingly resembles that of the upper band in the pure Kitaev model.

As further evidence, the fact that $M_2$ survives above $T_N$, even while $M_1$ is completely washed out indicates that the $M_2$ mode is not directly connected to the existence of long range magnetic order. As is observed in coupled S=1/2 spin chains[6] the existence of long-range magnetic order need not interfere considerably with the signatures of fractionalization at energy scales above that set by the ordering temperature. In the strictly 2D Kitaev model there is no true phase transition from the QSL to the high temperature paramagnet[39]. It is reasonable to expect that the $M_2$ mode remains stable at temperatures below the energy scale of $K$.

Taken together, the qualitative features from a complete quantum calculation using a Majorana fermion treatment can successfully provide a broadly consistent account of the inelastic neutron scattering data. This makes $\alpha$-RuCl$_3$ a prime candidate for realizing Kitaev and QSL physics. Further support for the presence of Kitaev QSL physics in $\alpha$-RuCl$_3$ is seen in recent Raman scattering measurements[22] which show broad response similar to that calculated for the pure Kitaev model[16] with a value of $K$ similar to that derived from neutron scattering. The prospect of finding fractionalized excitations in this class of magnets is a strong motivation for more detailed studies, in particular INS in single crystals of $\alpha$-RuCl$_3$, iridates, and related compounds, some of which are 3D[40,41]. Looking forward, it will also be of great interest to systematically investigate the effects of disorder and doping in these materials[42].

**Methods**

Commercial-RuCl$_3$ powder was purified in-house to a mixture of $\alpha$-RuCl$_3$ and $\beta$-RuCl$_3$, and converted to 99.9% phase pure $\alpha$-RuCl$_3$ by annealing at 500 °C. Single crystals of $\alpha$-RuCl$_3$ were grown using vapor transport with TeCl$_4$ as the transport agent. Samples were characterized by standard bulk techniques as well as both x-ray and neutron diffraction. The structure was consistent with the trigonal space group P3$_1$12 (#151) with room-temperature lattice constants a=b=5.97 Å, c=17.2 Å. Magnetic properties were measured with a Quantum Design (QD) Magnetic Property Measurement System in the temperature interval 1.8 K ≤ T ≤ 300 K. Temperature-dependent specific heat data were collected using a 14 T QD Physical Property Measurement System (PPMS) in the temperature range from 1.9 to 200 K. Our measurements of the specific heat and susceptibility are consistent with existing literature[20,23,24,27]. The magnetic susceptibility of powders fits a Curie-Weiss law over the range 200-300 K, with a temperature intercept of $\theta \approx 32$ K and a single-ion Ru effective moment of 2.2 $\mu_B$. Magnetic order appears for T $\leq$ 15 K leading to a broad specific heat anomaly. The detailed specific heat of single crystal specimens is sample dependent, but consistent with other groups[23,24,27], shows the onset of a broad anomaly near 14 K, and a sharper peak near 8 K, possibly with additional structure in between those temperatures. This complicated behavior is a consequence of stacking faults (see main text).

Powder x-ray measurements used a Panalytical Empyrian diffractometer employing Cu K$_\alpha$ radiation. Neutron diffraction data for structural refinement on a 5 gram powder sample of $\alpha$-RuCl$_3$ were collected at the POWGEN beamline at the Spallation Neutron Source, ORNL. The sample was loaded in a vanadium sample can under helium, and measured at T ≈ 10 K.

Neutron diffraction measurements to characterize the magnetic peaks in both powder and single crystals was performed at the HB-1A Fixed Incident Energy (FIE-TAX, $E_i$ = 14.6 meV) beamline at the High-Flux Isotope Reactor.  For powder diffraction, 4.7 grams of powder were packed into a cylindrical aluminum canister. For single crystal diffraction, one ~0.7 x 1.0 cm$^2$, 22.5 mg crystal was attached to a flat aluminum shim using Cytop glue. It was then sealed with indium into an aluminum canister with helium exchange gas and then aligned and confirmed to be a single domain sample using neutrons. This was attached to the cold-finger of a 4 K displex for performing the temperature scans. Inelastic neutron scattering of powder α-RuCl$_3$ was performed using the SEQUOIA chopper spectrometer[43].  The sample (5.3 grams) was sealed at room temperature in a 5 x 5 x 0.2 cm$^3$ flat aluminum sample can using helium exchange gas for thermal contact.  This was mounted to the cold finger of a closed cycle helium refrigerator for temperature control.  Empty can measurements were performed under the same conditions as the sample measurements. All inelastic data has been normalized to the incident proton charge and have the empty can background subtracted. Measurements were made with $E_i$=8, 25, and 1500 meV for the neutron incident energies.  The $E_i$=8 and 25 meV measurements were performed using the fine resolution 100 meV Fermi chopper slit package spinning at 180 Hz and the $T_0$ chopper spinning at 30 Hz.  The $E_i$=1500 meV measurements used the 700 meV coarse resolution Fermi chopper spinning at 600 Hz and the $T_0$ chopper spinning at 180 Hz[44].  The $E_i$=1500 meV configuration yields a calculated full width at half maximum (FWHM) energy resolution of approximately 97 meV at 200 meV energy transfer.  The FWHM elastic energy resolution is calculated to be 0.19 and 0.64 meV for the $E_i$ = 8 and 25 meV configurations respectively.  Care was taken to minimize the exposure of the sample to air, and after every exposure the sample was pumped for at least 30 minutes to remove adsorbed moisture. Refinements of the structure utilized FULLPROF, and confirmed the purity of the powder sample. Spin-wave simulations were performed using SpinW codes[45] and used the nominal symmetric honeycomb structure for α-RuCl$_3$[19,20]. The Ru$^{3+}$ form factor utilized was interpolated using the results of relativistic Dirac-Slater wave functions[46].

**Acknowledgements:**

Research using ORNL's HFIR and SNS facilities was sponsored by the Scientific User Facilities Division, Office of Basic Energy Sciences, U.S. Department of Energy. The work at University of Tennessee was funded in part by the Gordon and Betty Moore Foundation's EPiQS Initiative through Grant GBMF4416 (D.G.M. and L.L.).  Some of the work at Oak Ridge National Laboratory was supported by the U.S. Department of Energy, Office of Basic Energy Sciences, Materials Sciences and Engineering Division. We thank B. Chakoumakos, J. Rau, S. Toth, and F. Ye for valuable discussions.   We thank Dr. P. Whitfield from POWGEN beamline and Dr. Z. Gai from CNMS facility for helping with neutron diffraction and magnetic susceptibility measurements.

**Author contributions:**

S.N., A.B. and D.G.M. conceived the project and the experiment.  C.B., L.L., A.B., J-Q.Y., Y.Y., and D.G.M. made the samples. J-Q.Y., L.L., A.B. and C.B. performed the bulk measurements, A.B., A.A., M.B.S., G.E.G, M.L. and S.N. performed INS measurements, A.B., S.N., C.B., M.L., and D.A.T. analyzed the data. A.B., M.L., S.B. and S.N. performed SWT simulations. J. K., S.B., D.K. and R.M. carried out QSL theory calculations.  A.B. and S.N. prepared the first draft, and all authors contributed to writing the manuscript.

15. Ye, F. *et al.*, Direct evidence of a zigzag spin-chain structure in the honeycomb lattice: A neutron and x-ray diffraction investigation of single-crystal $Na_2IrO_3$. *Phys. Rev. B* **85**, 180403(R) (2012).

16. Knolle, J., Chern, G.-W., Kovrizhin, D. L., Moessner, R. & Perkins, N. B., Raman Scattering Signatures of Kitaev Spin Liquids in $A_2IrO_3$ Iridates with A = Na or Li. *Phys. Rev. Lett.* **113**, 187201 (2014).

17. Gretarsson, H. *et al.*, Magnetic excitation spectrum of $Na_2IrO_3$ probed with resonant inelastic x-ray scattering. *Phys. Rev. B* **87**, 220407(R) (2013).

18. Figgis, B. N., Lewis, J., Mabbs, F. E. & Webb, G. A., Magnetic Properties of Some Iron(III) and Ruthenium(III) Low-spin Complexes. *J. Chem. Soc.* **A**, 422 - 426 (1966).

19. Fletcher, J. M. *et al.*, Anhydrous Ruthenium Chlorides. *Nature* **199**, 1089-1090 (1963).

20. Fletcher, J. M., Gardner, W. E., Fox, A. C. & Topping, G., X-Ray, Infrared, and Magnetic Studies of $\alpha$- and $\beta$-Ruthenium Trichloride. *Journal of Chem. Society* **A**, 1038-1045 (1967).

21. Plumb, K. W. *et al.*, α-$RuCl_3$: A spin-orbit assisted Mott insulator on a honeycomb lattice. *Phys. Rev. B* **90**, 041112(R) (2014).

22. Sandilands, L. J. *et al.*, Scattering Continuum and Possible Fractionalized Excitations in α-$RuCl_3$. *Phys. Rev. Lett.* **114**, 147201 (2015).

23. Sears, J. A. *et al.*, Magnetic order in α-$RuCl_3$: a honeycomb lattice quantum magnet with strong spin-orbit coupling, Available at http://arxiv.org/abs/1411.4610 (2014).

24. Shankar, V. V., Kim, H.-S. & Kee, H.-Y., Kitaev magnetism in honeycomb $RuCl_3$ with intermediate spin-orbit coupling, Available at http://arxiv.org/abs/1411.6623 (2014).

25. Majumder, M. *et al.*, Anisotropic $Ru^{3+}$ $4d^5$ magnetism in the α-$RuCl_3$ honeycomb system: susceptibility, specific heat and Zero field NMR, Available at http://arxiv.org/abs/1411.6515v1 (2014).

26. Sandilands, L. J. *et al.*, Orbital excitations in the 4d spin-orbit coupled Mott insulator α-RuCl3, Available at http://arxiv.org/abs/1503.07593 (2015).

27. Kubota, Y., Tanaka, H., Ono, T., Narumi, Y. & Kindo, K., Successive magnetic phase transitions in α-$RuCl_3$: XY-like frustrated magnet on the honeycomb lattice, Available at http://arxiv.org/abs/1503.03591v1 (2015).
11

**Figure Captions**

Fig. 1: (a)-(c) The structure of α-RuCl$_3$ (space group No. 151, P3$_1$12). (a) In-plane honeycomb structure showing edge sharing RuCl$_6$ octahedra and the unit cell of the honeycomb lattice. (b) View along the *c* axis showing the stacking of honeycomb layers in the unit cell, with Ru atoms in each layer denoted by the colors red, blue or green. The different intralayer Ru-Ru bonds, corresponding to the index "m" in equation (1), are labeled in the green layer as α, β, or γ, each with distance a/√3. The two-dimensional zig-zag magnetic structure is illustrated by the spins on the red layer. (c) Side view of the unit cell showing the offsets along the *c* axis. Values noted are for room temperature lattice constants. (d) Specific heat of powder α-RuCl$_3$. The solid red line is a fit of the data following the two dimensional Debye model
$C_p(T) = ANk \left(\frac{T}{\theta_D}\right)^2 \int_0^{\frac{\theta_D}{T}} \frac{x^2}{e^x - 1} dx$ for T > 16 K, and for T < 16 K an empirical function describing the anomaly associated with magnetic order. The inset in (d) shows a close-up of the anomaly associated with the low temperature magnetic ordering transition at T$_N$ ≈ 14 K in powder samples. (e) Order parameter plot of the (1/2 0 3/2) magnetic Bragg peak (Q= 0.81 Å$^{-1}$) in powder samples. The solid blue line is a power law fit to the data above 9 K yielding T$_N$ = 14.6(3) K, with β = 0.37(3). (f) Similar plot for single crystals showing two coexisting ordering wave vectors (1/2 0 1) with T$_{N1}$ = 7.6(2) K (green) and (1/2 0 3/2) with T$_{N2}$ = 14.2(8) K (blue). Note that the (1/2 0 1) peak loses intensity sharply, as compared to the (1/2 0 3/2). Inset: picture of the single crystal (22.5 mg) used in these measurements.

Fig. 2: Spin-orbit coupling mode in α-RuCl$_3$, measured at SEQUOIA with incident energy E$_i$ = 1.5 eV, and T = 5 K. (a) Difference between data integrated over the ranges Q = [2.5, 4.0] Å$^{-1}$ and [4.5, 6.0] Å$^{-1}$, subtracted point by point, illustrating the enhanced signal at low Q. The solid line is a fit to a background plus a Gaussian peak centered at 195 ± 11 meV with HWHM 48 ± 6 meV. With the settings used for the measurement the width is resolution limited. (b) Intensity for various values of wave-vector integrated over the energy range [150, 250] meV (each point represents a summation in Q over 0.5 Å$^{-1}$ except for the first point which is over 0.26 Å$^{-1}$). The solid line shows a two parameter fit of the data to the equation $A \cdot |f^{mag}(Q)|^2 + B$, where $f^{mag}(Q)$ is the Ru$^{3+}$ magnetic form factor in the spherical approximation. The shaded area represents the contribution arising from magnetic scattering. Inset: A schematic of the many-electron energy levels for $d^5$ electrons in the strong octahedral field (i.e. low spin) limit with spin-orbit coupling showing the J$_{1/2}$ to J$_{3/2}$ transition at energy 3λ/2.

Fig. 3: Collective magnetic excitations measured using time-of-flight INS at SEQUOIA on $\alpha$-RuCl$_3$ powder with incident energy $E_i$ = 25 meV. (a) False color plot of the data at T = 5 K showing magnetic modes ($M_1$ and $M_2$) with band centers near E = 4 and 6 meV. $M_1$ shows an apparent minimum near Q = 0.62 Å$^{-1}$, close to the magnitude of the M point of the honeycomb reciprocal lattice. The white arrow shows the concave lower edge of the $M_1$ mode. The yellow P denotes an emerging phonon. (b) The corresponding plot above $T_N$ at T = 15 K shows that $M_1$ has disappeared leaving strong quasi-elastic scattering at lower values of Q and E. (c) Constant-Q cuts through the scattering depicted in (a) and (b) centered at wave-vectors indicated by the dashed lines. The cuts A and C are summed over the range [0.5,0.8] Å$^{-1}$ which includes the M point of the 2D reciprocal lattice, while B and D span [1.0,1.5] Å$^{-1}$. The data from 2 – 8 meV in cut B is fit (solid blue line) to a pair of Gaussians yielding peak energies $E_1$ = 4.1(1) meV and $E_2$ = 6.5(1) meV. The solid lines through cuts A, C and D are guides to the eye. (d) Constant-E cuts integrated over the energy range [2.5,3.0] meV, at 4 K (E) and 15 K (F). See text for detail.

Fig. 4: (a) Spin wave simulation for K-H model with (K, J) = (7.0, -4.6) meV with a ZZ ground state. The lattice is the honeycomb plane appropriate for the P3$_1$12 space group. (b) The calculated powder averaged scattering including the magnetic form factor. The white arrow shows the concave nature of the lower boundary in (Q, E) space, similar to the data in Fig. 3 (a). (c) Cuts through the data of Fig. 3(a) integrated over 0.2 Å$^{-1}$ wide bands of wave-vector centered at the values shown. Lines are guides to the eye. Note that actual data includes a large elastic response from Bragg and incoherent scattering. (d) The same cuts, through the calculated scattering shown in Fig. 4 (b). Inset: Phase diagram of the KH model, after Ref. 32. The various phases are denoted by different colors: spin liquid (SL, blue), antiferromagnetic (AFM, light violet), stripy (ST, green), ferromagnetic (FM, orange), and zig-zag (ZZ, red). The red dots represent the two solutions for $\alpha$-RuCl$_3$ as determined by the zone center spin wave mode energies.

Fig. 5. (a) Scattering from mode $M_1$ measured at T = 5 K using $E_i$ = 8 meV. Lower panel shows constant energy cuts over the energy ranges shown, centered at the locations labeled (G,H) in the upper panel. The absence of structured scattering below 2 meV confirms the gap in the magnetic excitation spectrum. (b) A constant-Q cut (B in Fig. 3(c), with a linear background term subtracted) of the T= 5 K data at the location of $M_2$ is compared with the corresponding cut in the simulated spin-wave scattering from upper mode convolved with the instrumental resolution. The arrow marked "R" shows the FWHM of the resolution. (c) Constant-E cuts of the data through the $M_2$ mode above (red triangles) and below (blue squares) $T_N$. The lines are guides to the eye. (d) The powder average scattering calculated from a 2D isotropic Kitaev model, with antiferromagnetic K, using the results of Ref. 10, including the magnetic form factor. The upper feature is broad in energy and decreases in strength as Q increases.

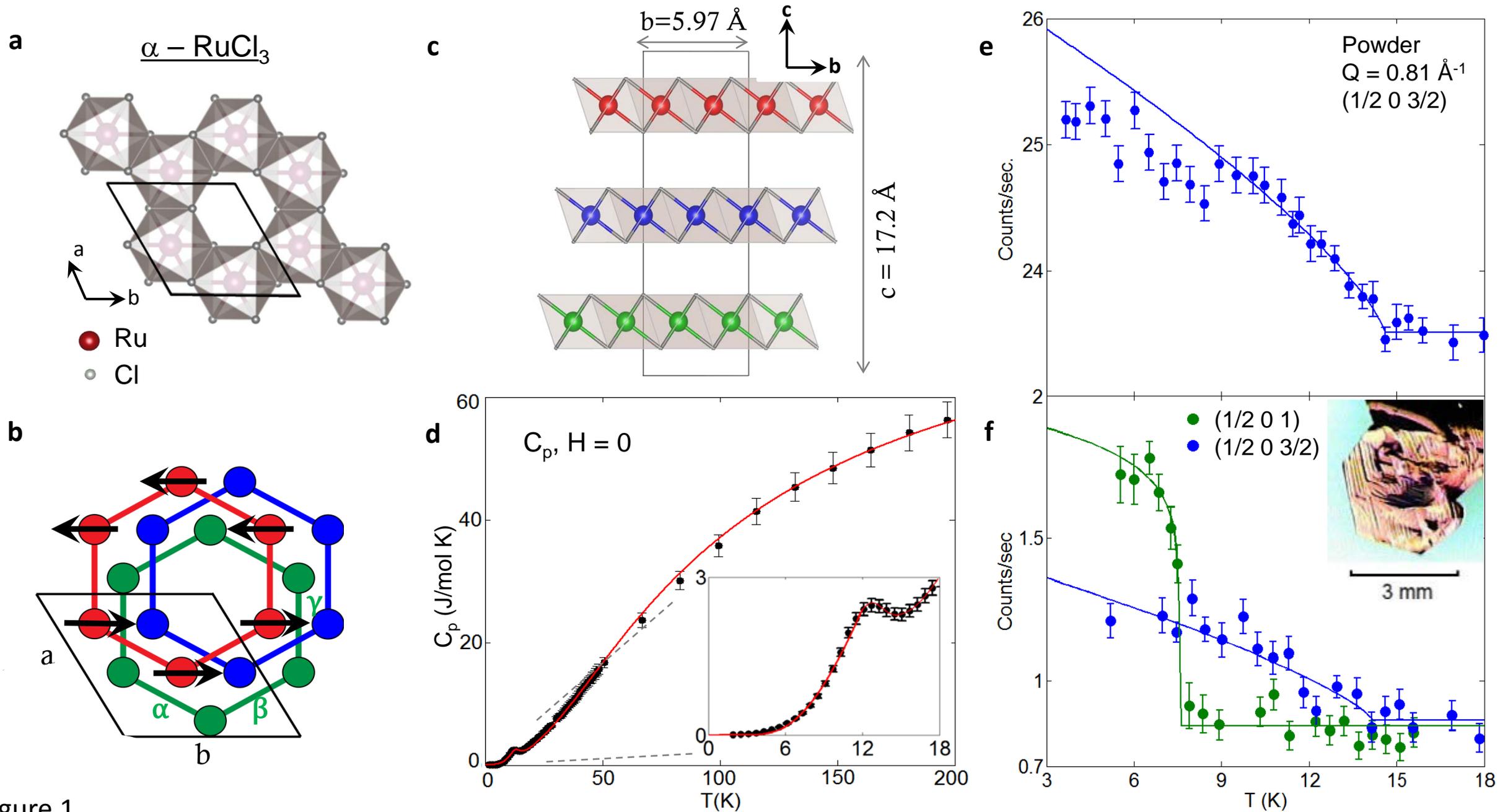

Figure 1

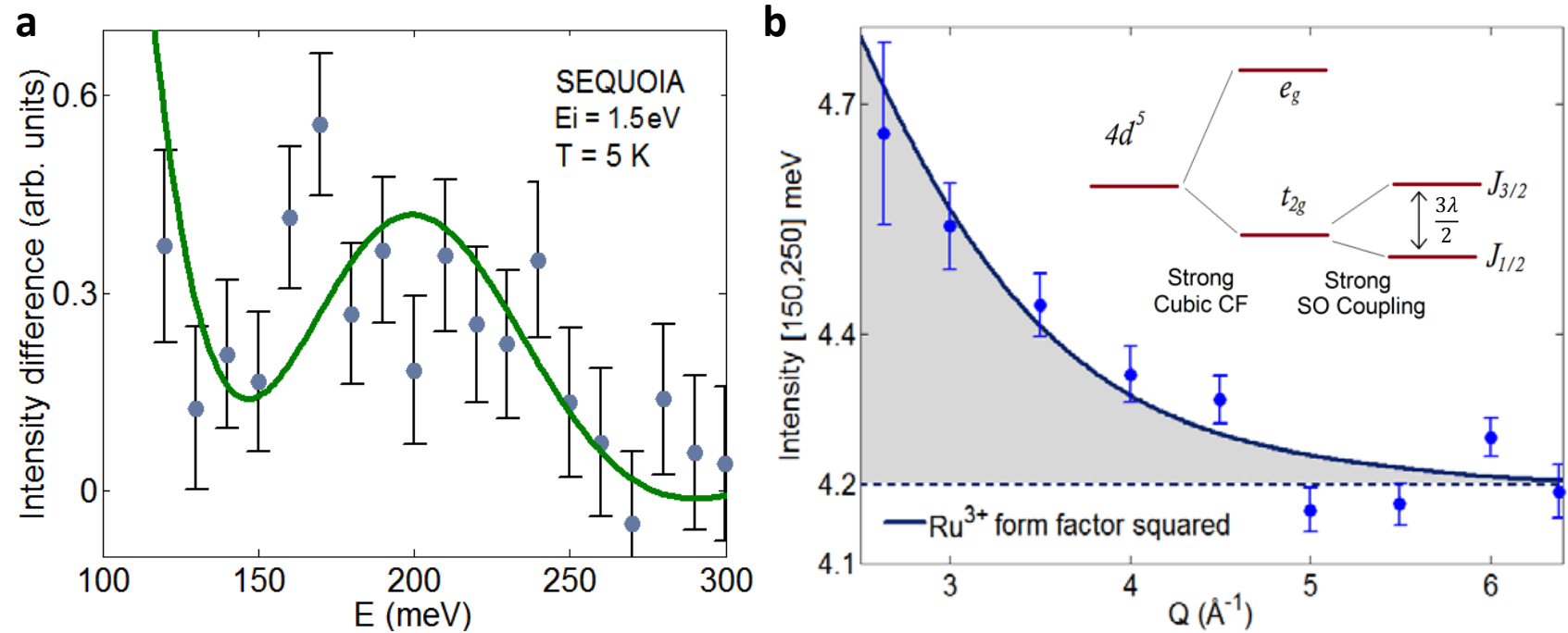

Figure 2

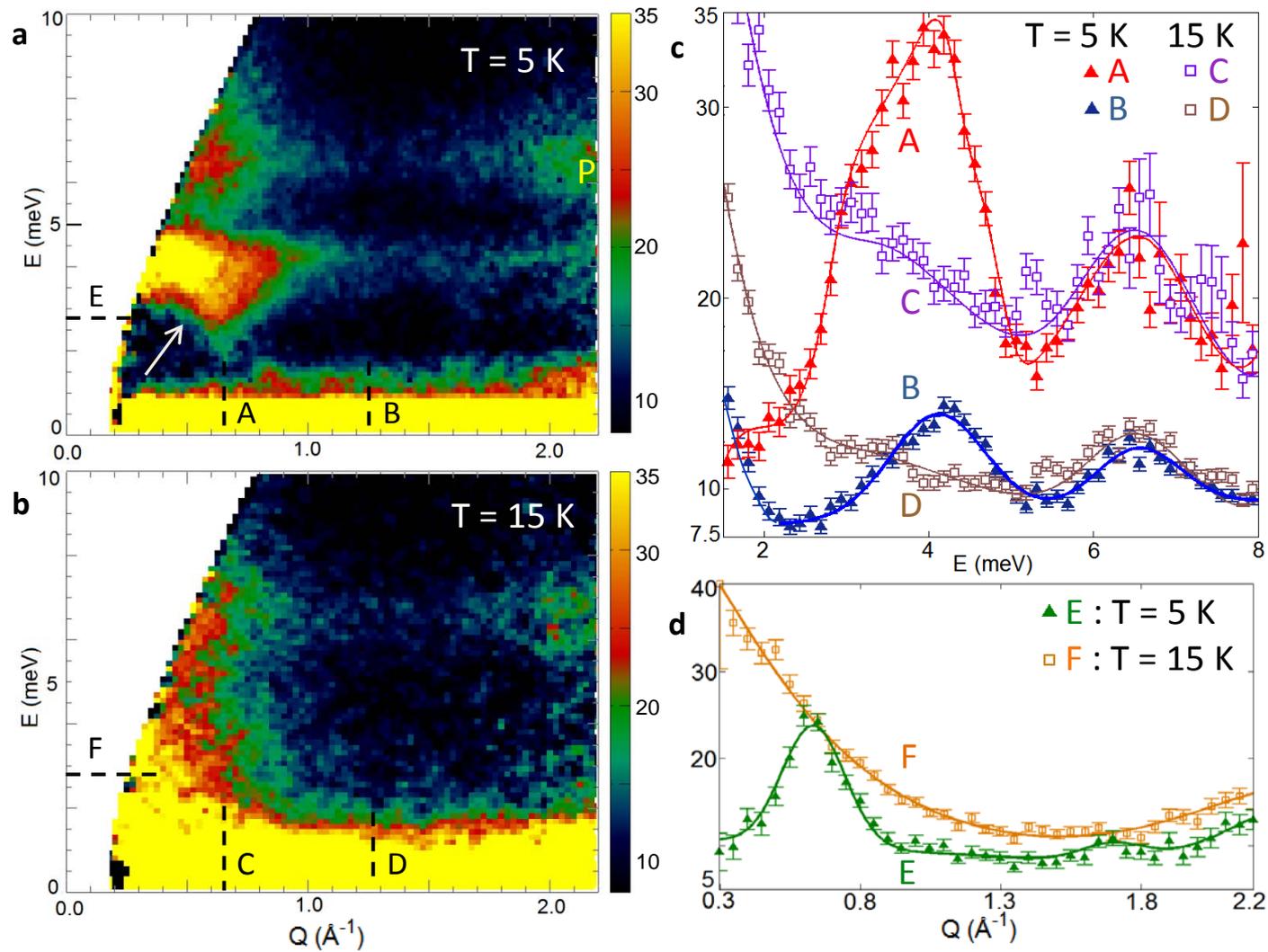

Figure 3

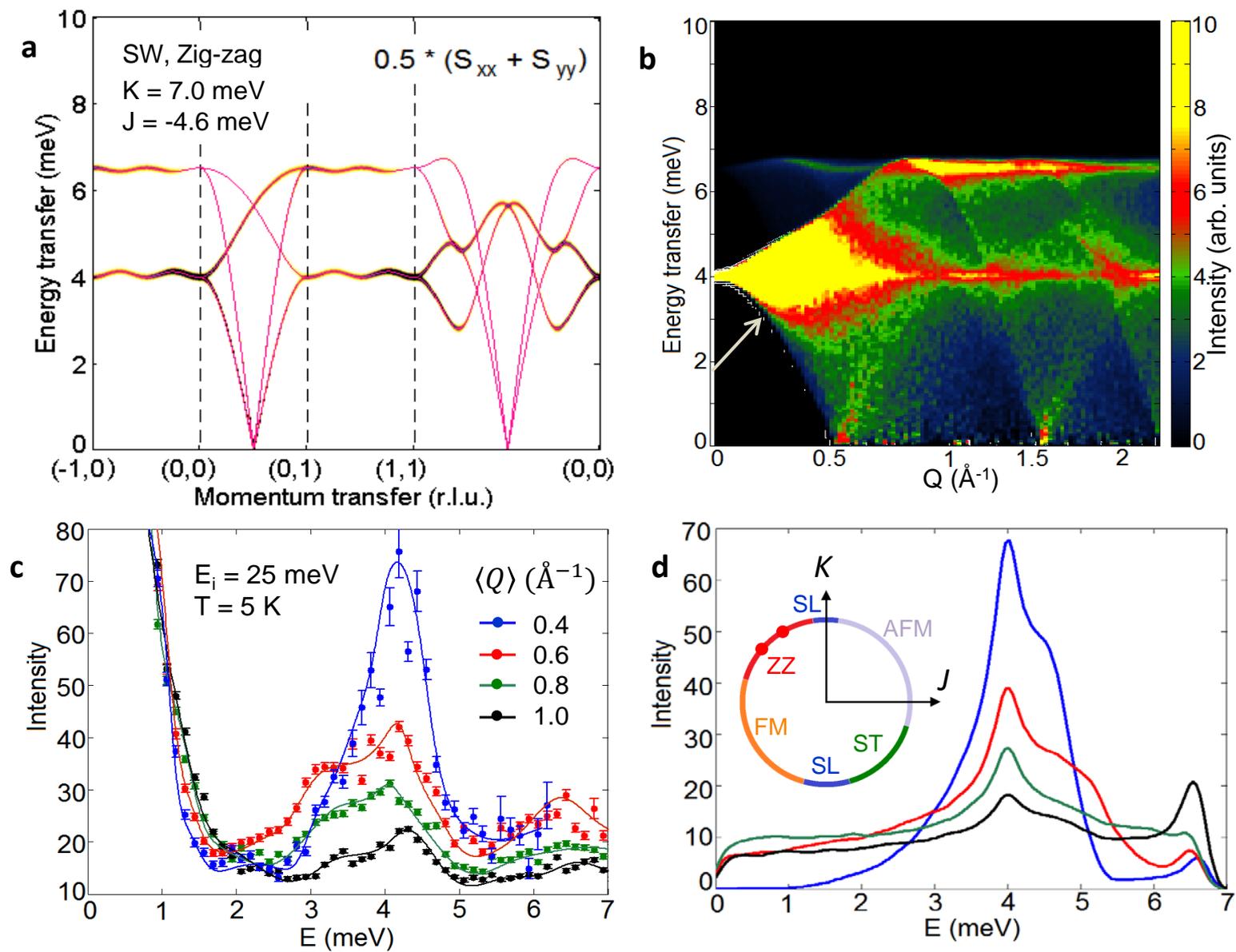

Figure 4

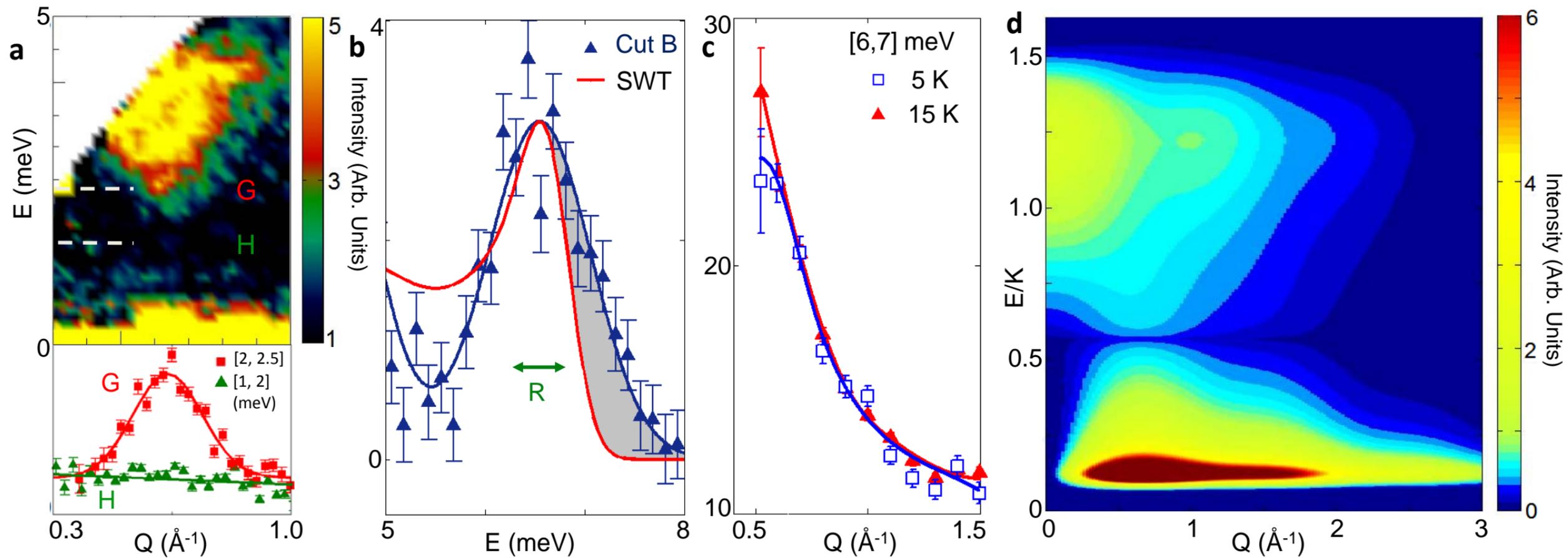

Figure 5